\documentclass[a4paper,11pt]{article}
\usepackage{pos}
\usepackage{subfigure}
\usepackage{orcidlink}

\title{Investigating Ultra-Low Energy Ionization Yield from Nuclear Recoils 
in Semiconductor Detectors via Molecular Dynamics Simulations}
\ShortTitle{Low-energy Quenching Factors for semiconductor detectors}

\author*[a]{Chang-Hao Fang\orcidlink{0000-0003-2474-906X}}
\onbehalf{on behalf of the CDEX collaboration}

\affiliation[a]{College of physics, Sichuan University\\
  24 South Section 1, 1st Ring Road, Chengdu, China}

\emailAdd{fangch@stu.scu.edu.cn}

\abstract{Nuclear recoil ionization yield constitutes a critical uncertainty 
source in low-energy detection for dark matter (DM) and coherent elastic 
neutrino-nucleus scattering (CE$\nu$NS) experiments. 
We present a novel methodology employing molecular dynamics simulations to 
assess ionization yields in crystalline semiconductor detectors. 
This non-parameterized approach resolving inherent limitations 
of traditional Lindhard model through explicit incorporation of crystal condensed 
matter effects, facilitating a seamless reliability from high-energy ($E>10$\,keV) 
to electron-hole pair (EHP) regimes. 
Our model achieves the best agreement with 
experimental data in silicon to date, especially at the minimal energy level of a 
single EHP.
Meticulously consideration of ion transport mechanisms reveals fundamental ionization 
yield distributions, superseding conventional single-value models. 
The distributional paradigm extends the DM-nucleon elastic scattering exclusion 
limit to 0.29\,GeV/$c^2$ under single-EHP sensitivity. 
We further report advancements in 
modeling quantum effects and channeling phenomena affecting ionization yields in 
high-purity germanium detectors.}

\FullConference{19th International Conference on Topics in Astroparticle and Underground Physics (TAUP2025)\\
24–30 Aug 2025\\
Xichang, China\\}

\begin{document}
\maketitle

\section{Introduction}
Coherent elastic neutrino-nucleus scattering (CE$\nu$NS)~\cite{freedmanCoherentEffectsWeak1974} and Weakly 
Interacting Massive Particles (WIMPs) interactions with nuclei~\cite{goodmanDetectabilityCertainDarkmatter1985} 
via the weak neutral current both produce 
substantial scattering cross sections, 
offering a powerful experimental probe of the Standard Model and potential new physics. 
Recent advances in semiconductor-based, low-threshold ionization detectors, particularly those capable of resolving 
individual electron-hole pairs~\cite{renDesignCharacterizationPhononmediated2021,tiffenbergSingleElectronSinglePhotonSensitivity2017},
have supported measurements of these low-momentum-transfer processes. 
In ionization detectors, the observable energy from a nuclear recoil is expressed as an equivalent 
electron energy \( E_\mathrm{ee} \), related to the original nuclear recoil energy \( E_\mathrm{nr} \) 
through the ionization yield (also known as the quenching factor, QF) via
\begin{equation}
  \label{eq:QF-def}
E_\mathrm{ee} = \mathrm{QF} \cdot E_\mathrm{nr}.
\end{equation}
To achieve the expected $E_\mathrm{ee}$ spectra, the expected $E_\mathrm{nr}$ spectra from neutrinos 
and dark matter must be corrected for the detector response, specifically, the QF. 
However, conventional models based on Lindhard theory exhibit significant discrepancies at low energies 
and fail to understand experimental measurements of the QF in this 
regime~\cite{lindhard1963integral,sorensenAtomicLimitsSearch2015,
  sarkisStudyIonizationEfficiency2020,sarkisIonizationEfficiencyNuclear2023}. 
This discrepancy represents a major source of systematic uncertainty in the interpretation of results 
from dark matter and CE$\nu$NS experiments~\cite{xuDetectionCalibrationLowEnergy2023}.
To elucidate the behavior of the QF at low energies, this work introduces a 
novel paradigm based on molecular dynamics (MD) simulations that overcomes the inadequacies of 
traditional QF models at low energies and yields results in excellent 
agreement with measurements~\cite{fangMolecularDynamicsSimulations2025}.

\section{Molecular Dynamic Approach}

\begin{figure}[h]
\centering
\includegraphics[width=\linewidth]{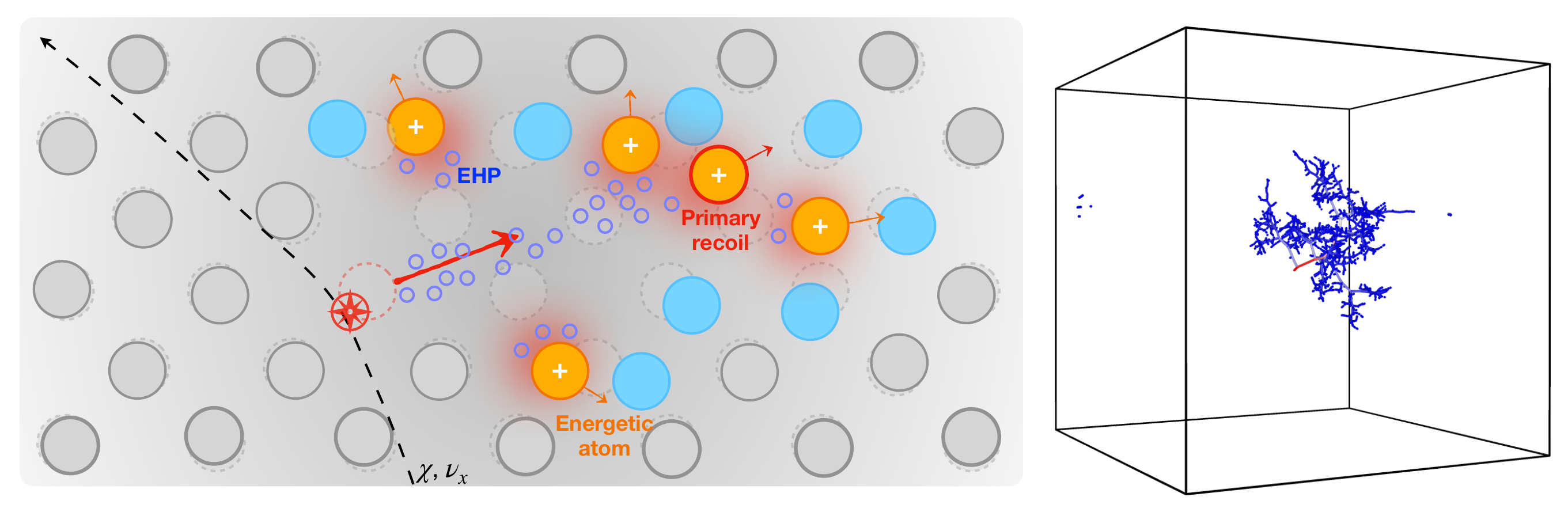}
\caption{
  \textbf{[left]} Ionization production scenario: primary recoil atom (red), displaced atoms (orange), and lattice-bound atoms (blue). 
Grey fog and shadows denote electronic gas and ionization, respectively; dashed circles show an ideal lattice without thermal relaxation.
\textbf{[right]} Energetic atom trajectories from a recoil event in MD simulation.\label{fig:qf-scenario}
}
\end{figure}

Following a nuclear recoil, the energy is deposited in the detector through two competing processes: nucleus-nucleus interactions 
and nucleus-electron interactions. 
As the recoiling nucleus propagates through the crystal lattice, it undergoes multiple scattering events with surrounding 
atoms (see Fig.~\ref{fig:qf-scenario}), dissipating energy into the nuclear subsystem and creating athermal phonons and lattice defects. 
Concurrently, inelastic interactions between the recoiling nucleus and electrons excite bound electrons, 
producing electron-hole pairs in semiconductors. 
This competition between energy-deposition processes for a finite recoil energy 
directly determines the ionization yield.

Our MD framework accurately captures the complex dynamics of nuclear recoils in semiconductor crystals. 
For silicon and germanium, we employ a combined potential approach: the Tersoff potential governs interactions at near-equilibrium distances 
($r \simeq 1\text{-}2\,\text{\AA}$), capturing the anisotropic covalent bonding in the diamond cubic structure~\cite{tersoffModelingSolidstateChemistry1989}, 
while the Ziegler-Biersack-Littmark (ZBL) potential describes close-encounter collisions at shorter ranges 
($r < 1\,\text{\AA}$) \cite{zieglerSRIMStoppingRange2015}. 
A spline-based interpolation is employed to ensure a smooth, 
physically consistent transition between the two regimes~\cite{devanathanDisplacementThresholdEnergies1998}.
Electronic energy loss is modeled as a frictional force proportional to the electronic stopping power $S_e$ from the 
validated SRIM software~\cite{zieglerSRIMStoppingRange2015, lohmannTrajectorydependentElectronicExcitations2020}. 
We implement these physics models in the \textsc{Lammps} MD simulation package~\cite{thompsonLAMMPSFlexibleSimulation2022}
to simulate the complete recoil cascade until energy deposition falls below the ionization threshold. 
The total energy deposited into the electronic system ($E_\mathrm{ee}$) is accumulated throughout the simulation, 
allowing direct calculation of the QF via Eq.~(\ref{eq:QF-def}).

\section{Results for Silicon and Germanium}
\subsection{Mean values of QF}
\begin{figure}[h]
  \centering
  \subfigure[]{
      \includegraphics[width=0.47\textwidth]{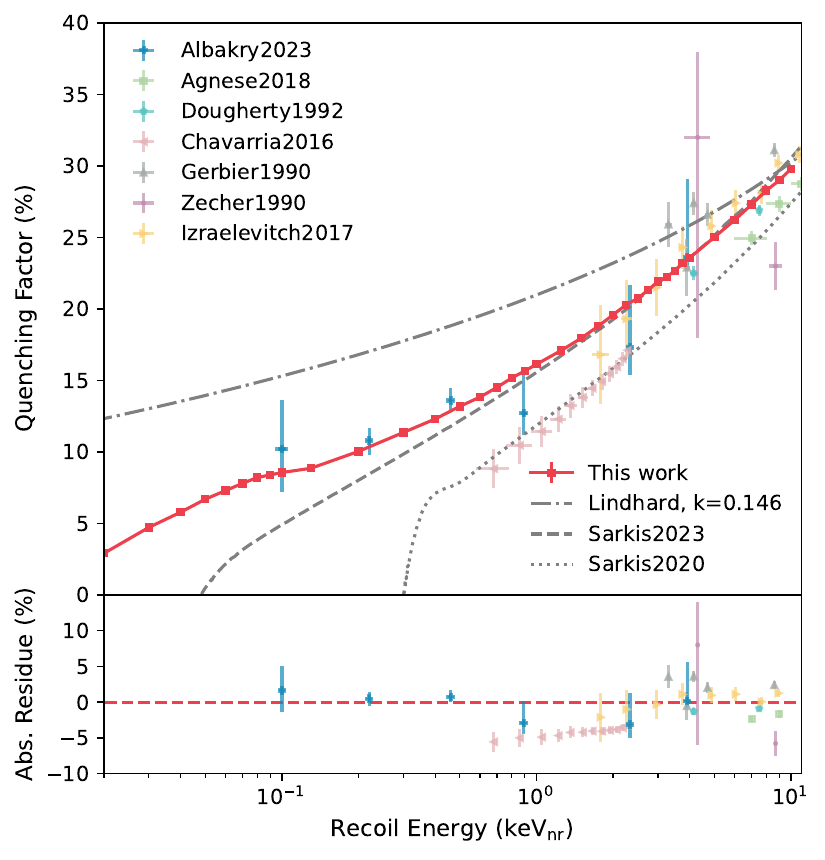}
      \label{subfig:Si-QF}
  }
  \subfigure[]{
      \includegraphics[width=0.47\textwidth]{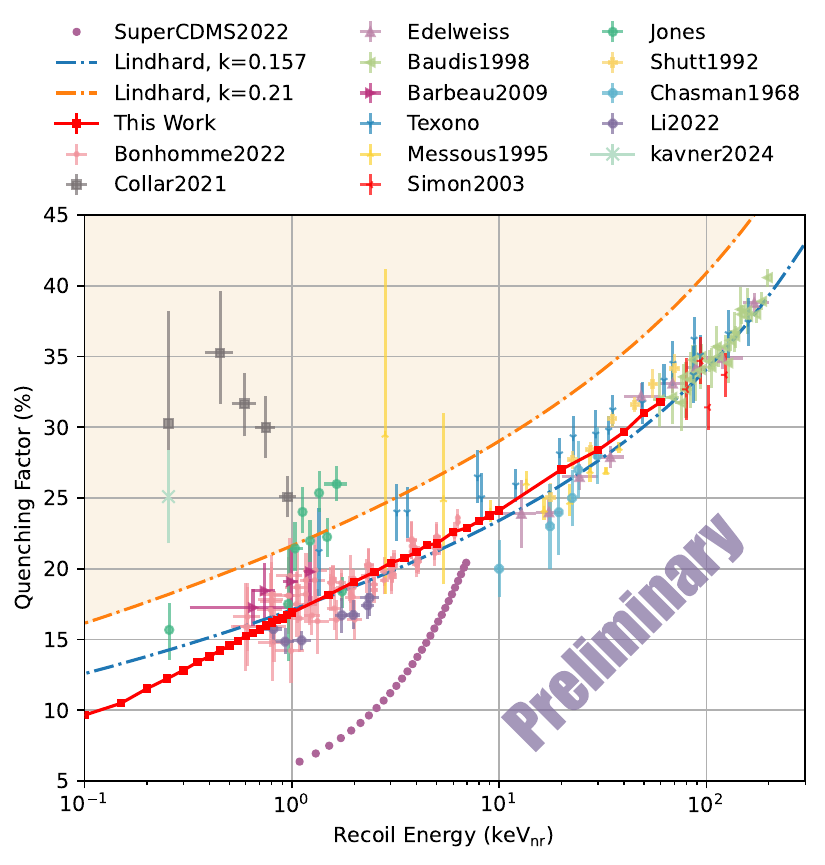}
      \label{subfig:Ge-QF}
  }
  \caption{Comparison of QFs from MD simulations (red dots) with experimental measurements (colored crosses) for 
  (a) silicon and (b) germanium detectors.\label{fig:QF-Exps}}
\end{figure}
Our MD simulations yield QFs across a wide range of recoil energies for silicon and germanium detectors. 
Figure~\ref{fig:QF-Exps} compares our results with existing experimental measurements. 
For silicon, our parameter-free MD-predicted QFs agree with low-energy measurements down to the electron-hole pairs level 
($100\,\mathrm{eV_{nr}}$)~\cite{albakryFirstMeasurementNuclearRecoil2023}, 
significantly outperforming the conventional Lindhard model, which overestimates the QF by more than 20\% below $4\,\mathrm{keV_{nr}}$.
Moreover, MD simulations reveal a non-trivial transition at $100\,\mathrm{eV_{nr}}$ arising from the directional dependence of the crystal lattice.
A detailed discussion of the physical reasons is given in Ref.~\cite{fangMolecularDynamicsSimulations2025}.
For germanium, preliminary results show good agreement with available measurements 
from $1\,\mathrm{keV_{nr}}$ to $60\,\mathrm{keV_{nr}}$. 
The behavior of the QF below about $1\,\mathrm{keV_{nr}}$ depends on the velocity threshold effect in the $S_e$ induced by the 
semiconductor bandgap~\cite{limElectronElevatorExcitations2016}.
It is currently being evaluated using time-dependent density functional theory (TDDFT) to obtain reliable low-energy QF results.
The MD method successfully captures the transition to lower QF values at energies approaching the lattice binding energy, where traditional models fail.

\subsection{Intrinsic Randomness and Distribution of QF}
\begin{figure}[h]
  \centering
      \subfigure[] {
          \includegraphics[width=0.47\textwidth]{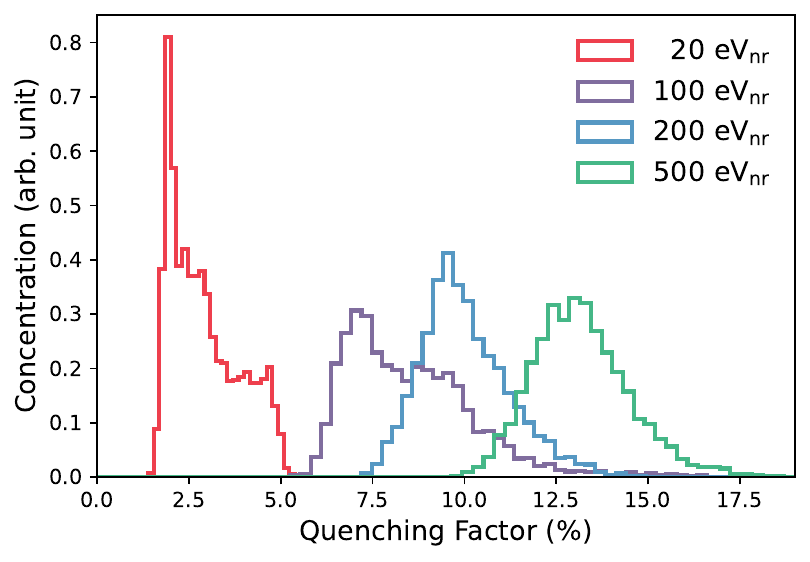}
          \label{subfig:QF-Dist-Low}
      }
      \subfigure[]{
          \includegraphics[width=0.47\textwidth]{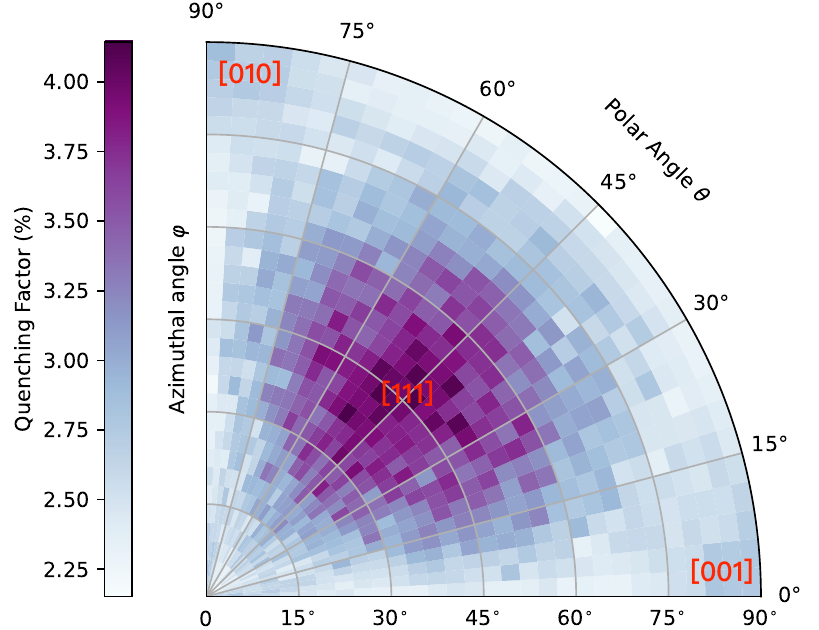}
          \label{subfig:QF-angle}
      }
  \caption{(a) Inherent distributions of sub-$\mathrm{keV_{nr}}$ QFs. (b) Angular dependence of QF at recoil energy of $20\,\mathrm{eV_{nr}}$.
}
  \label{fig:QF-Dist}
\end{figure}

Crucially, our simulations reveal that the QF varies significantly for a given recoil energy. 
Fig.~\ref{subfig:QF-Dist-Low} shows the random distribution of QF values in silicon across different recoil energies. 
At higher energies ($E_\mathrm{nr} > 200\,\mathrm{eV_{nr}}$), the distribution is relatively continuous, whereas below $100\,\mathrm{eV_{nr}}$, 
it becomes structured due to the crystal lattice. 
Fig.~\ref{subfig:QF-angle} demonstrates strong recoil-direction dependence at $20\,\mathrm{eV_{nr}}$, 
with QFs varying by up to a factor of 2 among different orientations. 
This directional dependence is most pronounced along the [111] crystal direction
and is primarily due to the highly anisotropic nature of crystal binding and the lattice structure,
as detailed in Ref.~\cite{fangMolecularDynamicsSimulations2025}.
Besides, calculations for both Si and Ge have revealed that high-energy secondary recoils generated in the cascade can lead to 
QFs exceeding 50\% due to channeling effects (which appear in the tail of the distribution), and related studies are currently underway.
These findings indicate that QF should be treated as a distribution rather than a deterministic function of recoil energy.
The broadening of the QFs brings a new perspective to longstanding dark matter and CE$\nu$NS analyses 
that have traditionally relied on a single, fixed QF value. 
The effects of these distributions deserve careful consideration in future work.

\section{Impact on Dark Matter Searches}

\begin{figure}
  \centering
  \includegraphics[width=0.6\textwidth]{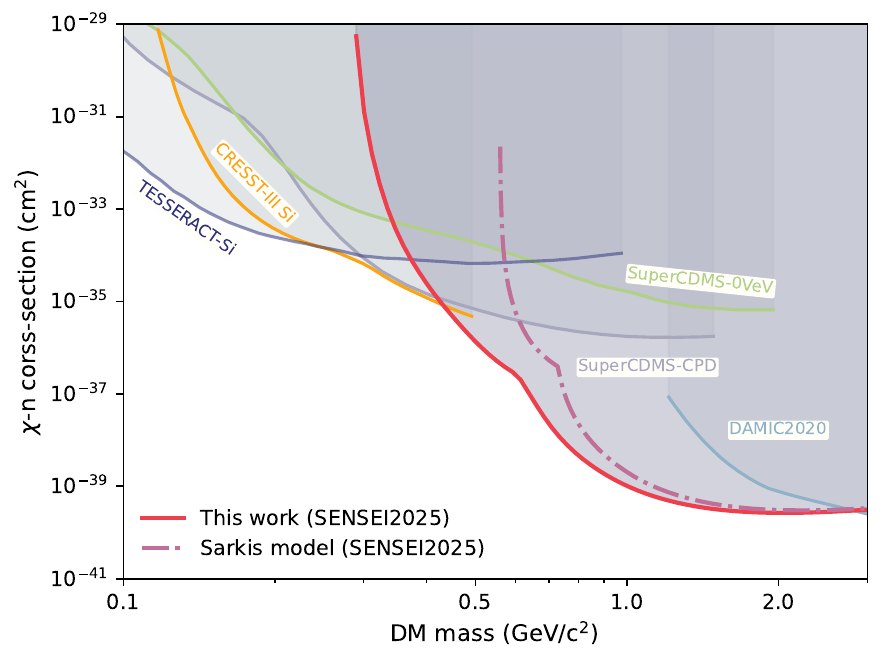}
  \caption{90\% C.L. upper limits for the $\chi$-N interaction. Red line is obtained from recent SENSEI 
  result~\cite{adariFirstDirectDetectionResults2025} using distribution perspective.
  \label{fig:DM-limit}
  }
\end{figure}

We evaluate the influence of QF models on the interpretation of dark matter
search results with the recent SENSEI result~\cite{adariFirstDirectDetectionResults2025}.
The expected event rate $\mathrm{d}R/\mathrm{d}E_\mathrm{nr}$ for
spin-independent dark matter-nucleon ($\chi$-N) couplings is derived using standard galactic halo
parameters in the elastic scattering model~\cite{lewinReviewMathematicsNumerical1996,
  baxterRecommendedConventionsReporting2021}.
Fig.~\ref{fig:DM-limit} illustrates the 90\% confidence level (C.L.) upper limit for the 
detection of $\chi$-N using Si detectors. Incorporating the perspective of QF distribution significantly enhances the sensitivity of 
$\chi$-N detection, and lowers the mass limit for $\chi$-N channel constraints to $0.29\,\mathrm{GeV/c^2}$. 
This advancement underscores the importance of considering QF distributions to improve detection 
capabilities and refine constraints on particle masses.

\section{Summary and Future Prospects}
Using MD as a framework to incorporate the properties of condensed-matter 
systems into ionization yield calculations has opened a new direction for 
evaluating low-energy ionization yields. 
This approach has achieved remarkable success in silicon detectors, 
even down to the level of electron-hole pairs. 
Calculations for germanium detectors are currently underway. 
Furthermore, this work reveals an intrinsic distributional broadening of the QF. 
In dark matter experiments, accounting for the QF distribution can significantly lower the detectable DM mass  
and enhance exclusion sensitivity.
QF broadening arising from the recoil cascade process is universal in ionization- and scintillation-based 
detection and warrants broad experimental attention.
Ongoing studies are investigating the impacts of semiconductor band gaps and channeling effects on the QF.

\acknowledgments 
This work was supported by the National Key Research and Development
Program of China (Contract No. 2023YFA1607103) and the National Natural
Science Foundation of China (Contracts No. 12441512, No. 11975159,
No. 11975162) provided support for this work.

\bibliography{ref.bib}

\begin{thebibliography}{20}%
\makeatletter
\providecommand \@ifxundefined [1]{%
 \@ifx{#1\undefined}
}%
\providecommand \@ifnum [1]{%
 \ifnum #1\expandafter \@firstoftwo
 \else \expandafter \@secondoftwo
 \fi
}%
\providecommand \@ifx [1]{%
 \ifx #1\expandafter \@firstoftwo
 \else \expandafter \@secondoftwo
 \fi
}%
\providecommand \natexlab [1]{#1}%
\providecommand \enquote  [1]{``#1''}%
\providecommand \bibnamefont  [1]{#1}%
\providecommand \bibfnamefont [1]{#1}%
\providecommand \citenamefont [1]{#1}%
\providecommand \href@noop [0]{\@secondoftwo}%
\providecommand \href [0]{\begingroup \@sanitize@url \@href}%
\providecommand \@href[1]{\@@startlink{#1}\@@href}%
\providecommand \@@href[1]{\endgroup#1\@@endlink}%
\providecommand \@sanitize@url [0]{\catcode `\\12\catcode `\$12\catcode `\&12\catcode `\#12\catcode `\^12\catcode `\_12\catcode `\%12\relax}%
\providecommand \@@startlink[1]{}%
\providecommand \@@endlink[0]{}%
\providecommand \url  [0]{\begingroup\@sanitize@url \@url }%
\providecommand \@url [1]{\endgroup\@href {#1}{\urlprefix }}%
\providecommand \urlprefix  [0]{URL }%
\providecommand \Eprint [0]{\href }%
\providecommand \doibase [0]{https://doi.org/}%
\providecommand \selectlanguage [0]{\@gobble}%
\providecommand \bibinfo  [0]{\@secondoftwo}%
\providecommand \bibfield  [0]{\@secondoftwo}%
\providecommand \translation [1]{[#1]}%
\providecommand \BibitemOpen [0]{}%
\providecommand \bibitemStop [0]{}%
\providecommand \bibitemNoStop [0]{.\EOS\space}%
\providecommand \EOS [0]{\spacefactor3000\relax}%
\providecommand \BibitemShut  [1]{\csname bibitem#1\endcsname}%
\let\auto@bib@innerbib\@empty
\bibitem [{\citenamefont {Freedman}(1974)}]{freedmanCoherentEffectsWeak1974}%
  \BibitemOpen
  \bibfield  {author} {\bibinfo {author} {\bibfnamefont {D.~Z.}\ \bibnamefont {Freedman}},\ }\href {https://doi.org/10.1103/PhysRevD.9.1389} {\bibfield  {journal} {\bibinfo  {journal} {Phys. Rev. D}\ }\textbf {\bibinfo {volume} {9}},\ \bibinfo {pages} {1389} (\bibinfo {year} {1974})}\BibitemShut {NoStop}%
\bibitem [{\citenamefont {Goodman}\ and\ \citenamefont {Witten}(1985)}]{goodmanDetectabilityCertainDarkmatter1985}%
  \BibitemOpen
  \bibfield  {author} {\bibinfo {author} {\bibfnamefont {M.~W.}\ \bibnamefont {Goodman}}\ and\ \bibinfo {author} {\bibfnamefont {E.}~\bibnamefont {Witten}},\ }\href {https://doi.org/10.1103/PhysRevD.31.3059} {\bibfield  {journal} {\bibinfo  {journal} {Phys. Rev. D}\ }\textbf {\bibinfo {volume} {31}},\ \bibinfo {pages} {3059} (\bibinfo {year} {1985})}\BibitemShut {NoStop}%
\bibitem [{\citenamefont {Ren}\ \emph {et~al.}(2021)\citenamefont {Ren}, \citenamefont {Bathurst}, \citenamefont {Chang} \emph {et~al.}}]{renDesignCharacterizationPhononmediated2021}%
  \BibitemOpen
  \bibfield  {author} {\bibinfo {author} {\bibfnamefont {R.}~\bibnamefont {Ren}}, \bibinfo {author} {\bibfnamefont {C.}~\bibnamefont {Bathurst}}, \bibinfo {author} {\bibfnamefont {Y.~Y.}\ \bibnamefont {Chang}}, \emph {et~al.},\ }\href {https://doi.org/10.1103/PhysRevD.104.032010} {\bibfield  {journal} {\bibinfo  {journal} {Phys. Rev. D}\ }\textbf {\bibinfo {volume} {104}},\ \bibinfo {pages} {032010} (\bibinfo {year} {2021})}\BibitemShut {NoStop}%
\bibitem [{\citenamefont {Tiffenberg}\ \emph {et~al.}(2017)\citenamefont {Tiffenberg}, \citenamefont {{Sofo-Haro}}, \citenamefont {{Drlica-Wagner}} \emph {et~al.}}]{tiffenbergSingleElectronSinglePhotonSensitivity2017}%
  \BibitemOpen
  \bibfield  {author} {\bibinfo {author} {\bibfnamefont {J.}~\bibnamefont {Tiffenberg}}, \bibinfo {author} {\bibfnamefont {M.}~\bibnamefont {{Sofo-Haro}}}, \bibinfo {author} {\bibfnamefont {A.}~\bibnamefont {{Drlica-Wagner}}}, \emph {et~al.},\ }\href {https://doi.org/10.1103/PhysRevLett.119.131802} {\bibfield  {journal} {\bibinfo  {journal} {Phys. Rev. Lett.}\ }\textbf {\bibinfo {volume} {119}},\ \bibinfo {pages} {131802} (\bibinfo {year} {2017})}\BibitemShut {NoStop}%
\bibitem [{\citenamefont {Lindhard}\ \emph {et~al.}(1963)\citenamefont {Lindhard}, \citenamefont {Nielsen}, \citenamefont {Scharff} \emph {et~al.}}]{lindhard1963integral}%
  \BibitemOpen
  \bibfield  {author} {\bibinfo {author} {\bibfnamefont {J.}~\bibnamefont {Lindhard}}, \bibinfo {author} {\bibfnamefont {V.}~\bibnamefont {Nielsen}}, \bibinfo {author} {\bibfnamefont {M.}~\bibnamefont {Scharff}}, \emph {et~al.},\ }\href {https://www.osti.gov/biblio/4701226} {\bibfield  {journal} {\bibinfo  {journal} {Mat. Fys. Medd. Dan. Vid. Selsk}\ }\textbf {\bibinfo {volume} {33}},\ \bibinfo {pages} {1} (\bibinfo {year} {1963})}\BibitemShut {NoStop}%
\bibitem [{\citenamefont {Sorensen}(2015)}]{sorensenAtomicLimitsSearch2015}%
  \BibitemOpen
  \bibfield  {author} {\bibinfo {author} {\bibfnamefont {P.}~\bibnamefont {Sorensen}},\ }\href {https://doi.org/10.1103/PhysRevD.91.083509} {\bibfield  {journal} {\bibinfo  {journal} {Phys. Rev. D}\ }\textbf {\bibinfo {volume} {91}},\ \bibinfo {pages} {083509} (\bibinfo {year} {2015})}\BibitemShut {NoStop}%
\bibitem [{\citenamefont {Sarkis}\ \emph {et~al.}(2020)\citenamefont {Sarkis}, \citenamefont {{Aguilar-Arevalo}},\ and\ \citenamefont {D'Olivo}}]{sarkisStudyIonizationEfficiency2020}%
  \BibitemOpen
  \bibfield  {author} {\bibinfo {author} {\bibfnamefont {Y.}~\bibnamefont {Sarkis}}, \bibinfo {author} {\bibfnamefont {A.}~\bibnamefont {{Aguilar-Arevalo}}},\ and\ \bibinfo {author} {\bibfnamefont {J.~C.}\ \bibnamefont {D'Olivo}},\ }\href {https://doi.org/10.1103/PhysRevD.101.102001} {\bibfield  {journal} {\bibinfo  {journal} {Phys. Rev. D}\ }\textbf {\bibinfo {volume} {101}},\ \bibinfo {pages} {102001} (\bibinfo {year} {2020})}\BibitemShut {NoStop}%
\bibitem [{\citenamefont {Sarkis}\ \emph {et~al.}(2023)\citenamefont {Sarkis}, \citenamefont {{Aguilar-Arevalo}},\ and\ \citenamefont {D'Olivo}}]{sarkisIonizationEfficiencyNuclear2023}%
  \BibitemOpen
  \bibfield  {author} {\bibinfo {author} {\bibfnamefont {Y.}~\bibnamefont {Sarkis}}, \bibinfo {author} {\bibfnamefont {A.}~\bibnamefont {{Aguilar-Arevalo}}},\ and\ \bibinfo {author} {\bibfnamefont {J.~C.}\ \bibnamefont {D'Olivo}},\ }\href {https://doi.org/10.1103/PhysRevA.107.062811} {\bibfield  {journal} {\bibinfo  {journal} {Phys. Rev. A}\ }\textbf {\bibinfo {volume} {107}},\ \bibinfo {pages} {062811} (\bibinfo {year} {2023})}\BibitemShut {NoStop}%
\bibitem [{\citenamefont {Xu}\ \emph {et~al.}(2023)\citenamefont {Xu}, \citenamefont {Barbeau},\ and\ \citenamefont {Hong}}]{xuDetectionCalibrationLowEnergy2023}%
  \BibitemOpen
  \bibfield  {author} {\bibinfo {author} {\bibfnamefont {J.}~\bibnamefont {Xu}}, \bibinfo {author} {\bibfnamefont {P.}~\bibnamefont {Barbeau}},\ and\ \bibinfo {author} {\bibfnamefont {Z.}~\bibnamefont {Hong}},\ }\href {https://doi.org/10.1146/annurev-nucl-111722-025122} {\bibfield  {journal} {\bibinfo  {journal} {Annu. Rev. Nucl. Part. Sci.}\ }\textbf {\bibinfo {volume} {73}},\ \bibinfo {pages} {95} (\bibinfo {year} {2023})}\BibitemShut {NoStop}%
\bibitem [{\citenamefont {Fang}\ \emph {et~al.}(2025)\citenamefont {Fang}, \citenamefont {Lin}, \citenamefont {Liu} \emph {et~al.}}]{fangMolecularDynamicsSimulations2025}%
  \BibitemOpen
  \bibfield  {author} {\bibinfo {author} {\bibfnamefont {C.-H.}\ \bibnamefont {Fang}}, \bibinfo {author} {\bibfnamefont {S.-T.}\ \bibnamefont {Lin}}, \bibinfo {author} {\bibfnamefont {S.-K.}\ \bibnamefont {Liu}}, \emph {et~al.},\ }\href {https://doi.org/10.1103/vhnm-356d} {\bibfield  {journal} {\bibinfo  {journal} {Phys. Rev. D}\ }\textbf {\bibinfo {volume} {112}},\ \bibinfo {pages} {L101303} (\bibinfo {year} {2025})}\BibitemShut {NoStop}%
\bibitem [{\citenamefont {Tersoff}(1989)}]{tersoffModelingSolidstateChemistry1989}%
  \BibitemOpen
  \bibfield  {author} {\bibinfo {author} {\bibfnamefont {J.}~\bibnamefont {Tersoff}},\ }\href {https://doi.org/10.1103/PhysRevB.39.5566} {\bibfield  {journal} {\bibinfo  {journal} {Phys. Rev. B}\ }\textbf {\bibinfo {volume} {39}},\ \bibinfo {pages} {5566} (\bibinfo {year} {1989})}\BibitemShut {NoStop}%
\bibitem [{\citenamefont {Ziegler}\ \emph {et~al.}(2015)\citenamefont {Ziegler}, \citenamefont {Biersack},\ and\ \citenamefont {Ziegler}}]{zieglerSRIMStoppingRange2015}%
  \BibitemOpen
  \bibfield  {author} {\bibinfo {author} {\bibfnamefont {J.~F.}\ \bibnamefont {Ziegler}}, \bibinfo {author} {\bibfnamefont {J.}~\bibnamefont {Biersack}},\ and\ \bibinfo {author} {\bibfnamefont {M.~D.}\ \bibnamefont {Ziegler}},\ }\href {http://www.srim.org} {\emph {\bibinfo {title} {{{SRIM}} - the Stopping and Range of Ions in Matter}}}\ (\bibinfo  {publisher} {SRIM},\ \bibinfo {address} {Chester, Maryland},\ \bibinfo {year} {2015})\BibitemShut {NoStop}%
\bibitem [{\citenamefont {Devanathan}\ \emph {et~al.}(1998)\citenamefont {Devanathan}, \citenamefont {Diaz De La~Rubia},\ and\ \citenamefont {Weber}}]{devanathanDisplacementThresholdEnergies1998}%
  \BibitemOpen
  \bibfield  {author} {\bibinfo {author} {\bibfnamefont {R.}~\bibnamefont {Devanathan}}, \bibinfo {author} {\bibfnamefont {T.}~\bibnamefont {Diaz De La~Rubia}},\ and\ \bibinfo {author} {\bibfnamefont {W.}~\bibnamefont {Weber}},\ }\href {https://doi.org/10.1016/S0022-3115(97)00304-8} {\bibfield  {journal} {\bibinfo  {journal} {Journal of Nuclear Materials}\ }\textbf {\bibinfo {volume} {253}},\ \bibinfo {pages} {47} (\bibinfo {year} {1998})}\BibitemShut {NoStop}%
\bibitem [{\citenamefont {Lohmann}\ \emph {et~al.}(2020)\citenamefont {Lohmann}, \citenamefont {Hole{\v n}{\'a}k},\ and\ \citenamefont {Primetzhofer}}]{lohmannTrajectorydependentElectronicExcitations2020}%
  \BibitemOpen
  \bibfield  {author} {\bibinfo {author} {\bibfnamefont {S.}~\bibnamefont {Lohmann}}, \bibinfo {author} {\bibfnamefont {R.}~\bibnamefont {Hole{\v n}{\'a}k}},\ and\ \bibinfo {author} {\bibfnamefont {D.}~\bibnamefont {Primetzhofer}},\ }\href {https://doi.org/10.1103/PhysRevA.102.062803} {\bibfield  {journal} {\bibinfo  {journal} {Phys. Rev. A}\ }\textbf {\bibinfo {volume} {102}},\ \bibinfo {pages} {062803} (\bibinfo {year} {2020})}\BibitemShut {NoStop}%
\bibitem [{\citenamefont {Thompson}\ \emph {et~al.}(2022)\citenamefont {Thompson}, \citenamefont {Aktulga}, \citenamefont {Berger} \emph {et~al.}}]{thompsonLAMMPSFlexibleSimulation2022}%
  \BibitemOpen
  \bibfield  {author} {\bibinfo {author} {\bibfnamefont {A.~P.}\ \bibnamefont {Thompson}}, \bibinfo {author} {\bibfnamefont {H.~M.}\ \bibnamefont {Aktulga}}, \bibinfo {author} {\bibfnamefont {R.}~\bibnamefont {Berger}}, \emph {et~al.},\ }\href {https://doi.org/10.1016/j.cpc.2021.108171} {\bibfield  {journal} {\bibinfo  {journal} {Computer Physics Communications}\ }\textbf {\bibinfo {volume} {271}},\ \bibinfo {pages} {108171} (\bibinfo {year} {2022})}\BibitemShut {NoStop}%
\bibitem [{\citenamefont {Albakry}\ \emph {et~al.}(2023)\citenamefont {Albakry}, \citenamefont {Alkhatib}, \citenamefont {Alonso} \emph {et~al.}}]{albakryFirstMeasurementNuclearRecoil2023}%
  \BibitemOpen
  \bibfield  {author} {\bibinfo {author} {\bibfnamefont {M.~F.}\ \bibnamefont {Albakry}}, \bibinfo {author} {\bibfnamefont {I.}~\bibnamefont {Alkhatib}}, \bibinfo {author} {\bibfnamefont {D.}~\bibnamefont {Alonso}}, \emph {et~al.},\ }\href {https://doi.org/10.1103/PhysRevLett.131.091801} {\bibfield  {journal} {\bibinfo  {journal} {Phys. Rev. Lett.}\ }\textbf {\bibinfo {volume} {131}},\ \bibinfo {pages} {091801} (\bibinfo {year} {2023})}\BibitemShut {NoStop}%
\bibitem [{\citenamefont {Lim}\ \emph {et~al.}(2016)\citenamefont {Lim}, \citenamefont {Foulkes}, \citenamefont {Horsfield} \emph {et~al.}}]{limElectronElevatorExcitations2016}%
  \BibitemOpen
  \bibfield  {author} {\bibinfo {author} {\bibfnamefont {A.}~\bibnamefont {Lim}}, \bibinfo {author} {\bibfnamefont {W.~M.~C.}\ \bibnamefont {Foulkes}}, \bibinfo {author} {\bibfnamefont {A.~P.}\ \bibnamefont {Horsfield}}, \emph {et~al.},\ }\href {https://doi.org/10.1103/PhysRevLett.116.043201} {\bibfield  {journal} {\bibinfo  {journal} {Phys. Rev. Lett.}\ }\textbf {\bibinfo {volume} {116}},\ \bibinfo {pages} {043201} (\bibinfo {year} {2016})}\BibitemShut {NoStop}%
\bibitem [{\citenamefont {Adari}\ \emph {et~al.}(2025)\citenamefont {Adari}, \citenamefont {Bloch}, \citenamefont {Botti} \emph {et~al.}}]{adariFirstDirectDetectionResults2025}%
  \BibitemOpen
  \bibfield  {author} {\bibinfo {author} {\bibfnamefont {P.}~\bibnamefont {Adari}}, \bibinfo {author} {\bibfnamefont {I.~M.}\ \bibnamefont {Bloch}}, \bibinfo {author} {\bibfnamefont {A.~M.}\ \bibnamefont {Botti}}, \emph {et~al.},\ }\href {https://doi.org/10.1103/PhysRevLett.134.011804} {\bibfield  {journal} {\bibinfo  {journal} {Phys. Rev. Lett.}\ }\textbf {\bibinfo {volume} {134}},\ \bibinfo {pages} {011804} (\bibinfo {year} {2025})}\BibitemShut {NoStop}%
\bibitem [{\citenamefont {Lewin}\ and\ \citenamefont {Smith}(1996)}]{lewinReviewMathematicsNumerical1996}%
  \BibitemOpen
  \bibfield  {author} {\bibinfo {author} {\bibfnamefont {J.}~\bibnamefont {Lewin}}\ and\ \bibinfo {author} {\bibfnamefont {P.}~\bibnamefont {Smith}},\ }\href {https://doi.org/10.1016/S0927-6505(96)00047-3} {\bibfield  {journal} {\bibinfo  {journal} {Astroparticle Physics}\ }\textbf {\bibinfo {volume} {6}},\ \bibinfo {pages} {87} (\bibinfo {year} {1996})}\BibitemShut {NoStop}%
\bibitem [{\citenamefont {Baxter}\ \emph {et~al.}(2021)\citenamefont {Baxter}, \citenamefont {Bloch}, \citenamefont {Bodnia} \emph {et~al.}}]{baxterRecommendedConventionsReporting2021}%
  \BibitemOpen
  \bibfield  {author} {\bibinfo {author} {\bibfnamefont {D.}~\bibnamefont {Baxter}}, \bibinfo {author} {\bibfnamefont {I.~M.}\ \bibnamefont {Bloch}}, \bibinfo {author} {\bibfnamefont {E.}~\bibnamefont {Bodnia}}, \emph {et~al.},\ }\href {https://doi.org/10.1140/epjc/s10052-021-09655-y} {\bibfield  {journal} {\bibinfo  {journal} {Eur. Phys. J. C}\ }\textbf {\bibinfo {volume} {81}},\ \bibinfo {pages} {907} (\bibinfo {year} {2021})}\BibitemShut {NoStop}%
\end{thebibliography}%

\end{document}